# Coherent manipulation of spin wave vector for polarization of photons in an atomic ensemble

*Shujing Li, Zhongxiao Xu, Haiyan Zheng, Xingbo Zhao, Yuelong Wu, Hai Wang\*, Changde Xie and Kunchi Peng*

*The State Key Laboratory of Quantum Optics and Quantum Optics Devices, Institute of Opto-Electronics, Shanxi University, Taiyuan, 030006, People's Republic of China*

We experimentally demonstrate the manipulation of two-orthogonal components of a spin wave in an atomic ensemble. Based on Raman two-photon transition and Larmor spin precession induced by magnetic field pulses, the coherent rotations between the two components of the spin wave is controllably achieved. Successively, the two manipulated spin-wave components are mapped into two orthogonal polarized optical emissions, respectively. By measuring Ramsey fringes of the retrieved optical signals, the $\pi/2$-pulse fidelity of ~96% is obtained. The presented manipulation scheme can be used to build an arbitrary rotation for qubit operations in quantum information processing based on atomic ensembles.

The coherent manipulation of quantum states in memory elements plays an important role in quantum information processing (QIP) [1-3]. Atomic ensembles with the ability of collectively enhanced coupling to a definite light mode [2] can serve as good quantum memory elements and have attracted considerable attention in recent years [1-3]. By using techniques of electromagnetically induced transparency (EIT), the storage and retrieval of photon states have been successfully accomplished with atomic ensembles [2]. The coherent memory time has reached to one second time scale in a BEC ensemble via controlled nonlinear interactions [4]. Quantum memories for single spin-wave excitation created via Raman scattering or EIT storage in atomic ensembles have been demonstrated [5-6]. Recently, the experimental studies on collective qubit memory used to achieve the atom-photon entanglement have had great progress. In these studies [7-9], two orthogonal [7] or two spatially distinct [8] spin waves, or two atomic ensembles shared a spin-wave excitation [9] are used to encode the long-time atomic qubit. The life time of qubit memory for optical lattice spin wave has reached to 3ms [7]. QIP schemes based-on atomic ensembles such as the entanglement swapping [1], the multipartite entanglement of atomic ensembles [3, 10], the controlled-NOT gate [10], and quantum computation with probabilistic quantum gates [11] have been proposed. In these proposed schemes, the memory qubits may be encoded in two orthogonal spin waves and single-bit operations are required. Single qubit gate operations are the $R(\theta, \varphi)$ and $R_z(\phi_z)$ rotations [12], which can be used to build an arbitrary rotation on the Bloch-sphere. The single qubit operations have been realized in single ion [12] or quantum dot system [13], while, it has not been demonstrated in atomic ensembles so far. The $R(\theta,\varphi)$ and $R_z(\phi_z)$ rotations in a spin-wave basis is a important precondition to perform a single qubit operation based on atomic ensemble. Recently, the experiment of the coherence transfer between two storage channels in BEC [4] has been realized. However, the required rotation of the spin wave for performing a single qubit operation still has not been achieved.

Here, we present the first experimental demonstration of the two unitary rotations $R(\theta,\varphi)$ and $R_z(\phi_z)$ in the basis formed by two-orthogonal components of a spin wave in an atomic ensemble. The spin wave is created by storing a left-circularly-polarized optical pulse in atomic ensemble via EIT dynamic scheme. Utilizing Raman two-photon transition, the $R(\theta,\varphi)$ rotation is achieved, with the coherence transfer efficiency of ~97%. Also, By applying a variable magnetic field pulse (with ~7μs width), the $R_z(\phi_z)$ rotation by any angle is also implemented. The π/2-pulse fidelity of about 96% is measured by observing Ramsey interference for a pair of π/2 Raman pulses. Such coherent manipulation can build an arbitrary rotation on the Bloch-sphere, thus it is an important progress toward implementing the memory qubit operation in atomic ensembles.

Fig.1(a), (b) and (c) illustrate the $^{87}$Rb relevant levels involved in the storage, Raman manipulation and readout processes, where $|\uparrow\rangle = |5^2S_{1/2}, F=1, M_F=+1\rangle$, $|\downarrow\rangle = |5^2S_{1/2}, F=1, M_F=-1\rangle$, $|s\rangle = |5^2S_{1/2}, F=2, M_F=+1\rangle$ and $|e\rangle = |5^2P_{1/2}, F'=1, M_F=0\rangle$. Considering the atomic spin wave associated with the coherence between the state $|s\rangle$ and the superposition state $|\Phi\rangle = \alpha|\uparrow\rangle + e^{-i\Delta}\beta|\downarrow\rangle$ ($|\alpha|^2 + |\beta|^2 = 1$), we introduce a collective, slowly varying atomic operator, appropriately averaged over a small but macroscopic volume with atomic numbers of $N_z \gg 1$ at positions $z$ [14], $\tilde{\rho}_{s\Phi}(z,t) = (N_z)^{-1}\sum_{j=1}^{N_z}\hat{\sigma}_{s\Phi}^j e^{i\omega_a t}$, $\hat{\sigma}_{s\Phi}^j = |s_j\rangle\langle\Phi_j|$ is the spin flip operator for the $j$th atom, $\omega_a$ is the frequency of the transition $|s\rangle \leftrightarrow |\uparrow\rangle$. The dark state polariton (DSP) describing the transfer between the spin wave and the optical signal field $\hat{\varepsilon}(z,t)$ can be written as [14]: $\hat{\Psi}(z,t) = \cos\vartheta(t)\hat{\varepsilon}(z,t) + \sin\vartheta(t)\sqrt{N}\tilde{\rho}_{s\Phi}(z,t)$, where $\tan\vartheta(t) = \sqrt{N}g/\Omega_C$, $\Omega_C$ is the Rabi frequency of the (reading or writing) coupling beam, $N$ is the atomic number. All the number states created by $\hat{\Psi}_k^+(z,t)$ are dark states:

$|D_n^k\rangle = (\sqrt{n!})^{-1}(\hat{\Psi}_k^+)^n|0\rangle|\Phi_1...\Phi_N\rangle$, where $\hat{\Psi}_k(t)$ is the plane-wave decomposition of $\hat{\Psi}(z,t) = \sum_k \hat{\Psi}_k(t)e^{ikz}$ [14]. Introducing $|\Phi_j\rangle = \alpha|\uparrow_j\rangle + e^{-i\Delta}\beta|\downarrow_j\rangle$ into the expression of $\tilde{\rho}_{s\Phi}(z,t)$, we have $\tilde{\rho}_{s\Phi}(z,t) = \alpha\tilde{\rho}_{s\uparrow}(z,t) + e^{-i\Delta}\beta\tilde{\rho}_{s\downarrow}(z,t)$, where $\tilde{\rho}_{s\uparrow}(z,t) = N_z^{-1}\sum_{j=1}^{N_z}|s_j\rangle\langle\uparrow_j|e^{i\omega_a t}$ and $\tilde{\rho}_{s\downarrow}(z,t) = N_z^{-1}\sum_{j=1}^{N_z}|s_j\rangle\langle\downarrow_j|e^{i\omega_a t}$ are the two spin-coherence operators, the real number $\alpha$ ($\alpha^2$) and $\beta$ ($\beta^2$) can be regarded as coherence amplitudes (populations) of the $\tilde{\rho}_{s\uparrow}(z,t)$ and $\tilde{\rho}_{s\downarrow}(z,t)$ components respectively, $\Delta$ is the relative phase between them. Since the spin-coherence operators $\tilde{\rho}_{s\uparrow}(z,t)$ and $\tilde{\rho}_{s\downarrow}(z,t)$ are orthogonal to each other ($\tilde{\rho}_{s\uparrow}(z,t)\tilde{\rho}_{s\downarrow}^+(z,t) = 0$), they form a spin-coherence basis $\{|s\rangle\langle\uparrow|$ and $|s\rangle\langle\downarrow|\}$. In this basis, the spin wave can be rewritten as a vector: $\hat{S}_{s\Phi}(z,t) = \langle|\tilde{\rho}_{s\Phi}(z,t)|\rangle\hat{\sigma}_{s\Phi} = \langle|\tilde{\rho}_{s\Phi}(z,t)|\rangle(\alpha\hat{\sigma}_{s\uparrow} + e^{-i\Delta}\beta\hat{\sigma}_{s\downarrow})$, where $\langle|\tilde{\rho}_{s\Phi}(z,t)|\rangle$ is the expectation value of the spin wave, $\hat{\sigma}_{sl} = (N_z)^{-1}\sum_{j=1}^{N_z}|s_j\rangle\langle l_j|$ ($l = \uparrow, \downarrow$) is the basis vector. With the projections of the basis vector $\langle s|\hat{\sigma}_{s\uparrow}|\uparrow\rangle = \langle s|\hat{\sigma}_{s\downarrow}|\downarrow\rangle = 1$ (where $\langle l| = \sum_{j=1}^{N_z}\langle l_j|$, $l = \uparrow, \downarrow$ and s), we calculate the projections of spin wave on the basis vectors: $\langle s|\hat{S}(z,t)|\uparrow\rangle = \alpha\langle|\tilde{\rho}_{s\Phi}(z,t)|\rangle$, $\langle s|\hat{S}(z,t)|\downarrow\rangle = e^{-i\Delta}\beta\langle|\tilde{\rho}_{s\Phi}(z,t)|\rangle$. Writing the basis unit vectors as: $\hat{\sigma}_{s\uparrow} = \begin{pmatrix}1\\0\end{pmatrix}$ and $\hat{\sigma}_{s\downarrow} = \begin{pmatrix}0\\1\end{pmatrix}$, the spin wave vector can be rewritten as: $\hat{S}_{s\Phi}(z,t) = \langle|\tilde{\rho}_{s\Phi}(z,t)|\rangle\begin{pmatrix}\alpha\\e^{-i\Delta}\beta\end{pmatrix}$, which can be regarded as the superposition of the two-orthogonal spin-wave components $\hat{S}_{s\uparrow}(z,t) = \langle|\tilde{\rho}_{s\Phi}(z,t)|\rangle\begin{pmatrix}\alpha\\0\end{pmatrix}$ and $\hat{S}_{s\downarrow}(z,t) = \langle|\tilde{\rho}_{s\Phi}(z,t)|\rangle\begin{pmatrix}0\\\beta\end{pmatrix}$. $\hat{S}_{s\uparrow}(z,t)$ and $\hat{S}_{s\downarrow}(z,t)$ will be converted to left- and right-circularly ($\sigma^+$- and $\sigma^-$-) polarized optical fields $\mathcal{E}_\pm^{out}$, respectively, when a reading beam is used for reading the spin wave. Considering the retrieval efficiency, the retrieved optical field is: $\varepsilon^{out}(z,t) = \begin{pmatrix} \varepsilon_+^{out}(z,t) \\ \varepsilon_-^{out}(z,t) \end{pmatrix} = \sqrt{N} \langle|\tilde{\rho}_{s\Phi}(z,t)|\rangle \begin{pmatrix} \sqrt{\eta_\uparrow}\alpha \\ \sqrt{\eta_\downarrow}e^{-i\Delta}\beta \end{pmatrix}$, where $\eta_\uparrow$ ($\eta_\downarrow$) is the retrieval efficiency from the channel $\tilde{S}_{s\uparrow}(z,t)$ ($\tilde{S}_{s\downarrow}(z,t)$). The retrieval efficiency $\eta_\uparrow$ ($\eta_\downarrow$) is proportional to the OD (optical depth) [15] of the transition $|\uparrow\rangle \leftrightarrow |e\rangle$ ($|\downarrow\rangle \leftrightarrow |e\rangle$). Since the transitions $|\uparrow\rangle \leftrightarrow |e\rangle$ and $|\downarrow\rangle \leftrightarrow |e\rangle$ are symmetric, so $\eta_\uparrow = \eta_\downarrow = \eta$. Thus we have: $\begin{pmatrix} \varepsilon_+^{out}(z,t) \\ \varepsilon_-^{out}(z,t) \end{pmatrix} \propto \sqrt{N\eta} \langle|\tilde{\rho}_{s\Phi}(z,t)|\rangle \begin{pmatrix} \alpha \\ e^{-i\Delta}\beta \end{pmatrix}$. It is obvious that the readout field $\varepsilon^{out}(z,t)$ directly copy the polarization information of the spin wave vector.

We now explain why the $R(\theta,\varphi)$ and $R_z(\Delta\phi_z)$ rotation of the spin wave vector $\hat{S}(z,t)$ on the Bloch sphere can be implemented by means of Raman two-photon transition and controllable Larmor spin precession. For the initial condition of $\Phi(0) = |\uparrow\rangle$, the spin wave is expressed by $\hat{S}(z,0) = \langle|\tilde{\rho}_{s\uparrow}(z,0)|\rangle \begin{pmatrix} 1 \\ 0 \end{pmatrix}$. We will see that the coherence population can be transferred between $\hat{S}_{s\uparrow}(z,t)$ and $\hat{S}_{s\downarrow}(z,t)$ channels via Raman two-photon transition. The Raman laser beam consists of $\sigma^+$- and $\sigma^-$- polarized light fields $E_{R+}$ and $E_{R-}$, which couple to $|\uparrow\rangle \leftrightarrow |e\rangle$ and $|\downarrow\rangle \leftrightarrow |e\rangle$ transitions, respectively, with a two-photon detuning $\delta_R \approx 0$ and a larger single-photon detuning $\Delta_R$ (see Fig.1(b)). If both the nature width $\Gamma$ and the Rabi frequencies $\Omega_{R\pm} = \mu_\pm E_{R\pm}/h$ are significantly smaller than the detuning $\Delta_R$, the upper state $|e\rangle$ is adiabatically eliminated, and the effective interaction Hamiltonian is given by $\hat{H}_{eff} = -h/2 \sum_{j=1}^{N} (\Omega_R |\uparrow\rangle_j \langle\downarrow|_j + H.c)$, where $\Omega_R = \Omega_{R+}\Omega_{R-}/2\Delta_R$ is the Raman-Rabi frequency. From the Hamiltonian $\hat{H}_{eff}$, we obtain the solutions of the superposition state $\Phi(t)$ for an any atom and then solve the evolution equation of the spin wave $\hat{S}(z,t)$, which can be described by a unitary Raman rotation $R(\theta,\varphi)$ and we have $R(\theta,\varphi)\widetilde{S}(z,0) \to \widetilde{S}(z,t)$, where $R(\theta,\varphi) = \begin{pmatrix} \cos\frac{\theta}{2} & -ie^{i\varphi}\sin\frac{\theta}{2} \\ -ie^{-i\varphi}\sin\frac{\theta}{2} & \cos\frac{\theta}{2} \end{pmatrix}$, $\theta = \Omega_R t$ and $\varphi = \varphi_+ - \varphi_-$ is the phase difference between $\sigma^+$ and $\sigma^-$ Raman fields $E_{R+}$ and $E_{R-}$. In the following we will discuss why the relative phase $\Delta$ can be controllably changed by Larmor precession. In a magnetic field, the two spin-wave components $\hat{S}_{s\uparrow}(z,t)$ and $\hat{S}_{s\downarrow}(z,t)$ experience Larmor precession [16] and get phase shifts $\psi_1$ and $\psi_2$ respectively, which introduce a relative phase $\phi_z = \psi_1 - \psi_2$ after an evolution time interval $t$. The evolution of the spin wave is described by $\hat{S}(z,t) = R_z(\phi_z)\hat{S}(z,0) = \langle|\tilde{\rho}_{s\uparrow}(z,0)|\rangle\begin{pmatrix}\alpha \\ e^{i\phi_z}\beta\end{pmatrix}$, where $R_z(\phi_z) = \begin{pmatrix} 1 & 0 \\ 0 & e^{i\phi_z} \end{pmatrix}$ is the matrix rotation and the initial spin wave is assumed to be $\hat{S}(z,0) = \langle|\tilde{\rho}_{s\uparrow}(z,0)|\rangle\begin{pmatrix}\alpha \\ \beta\end{pmatrix}$. It is obvious that the spin wave acquire a relative phase $\Delta = \phi_z$ due to Larmor precession.

The experimental setup is shown in Fig.1(d). A cold $^{87}$Rb atomic cloud released from a magneto-optical trap (MOT) serves as the atomic ensemble. The measured optical depth of the cold atoms is about 1.5 and the trap temperature can reach ~200 μK. A 780 nm $\sigma^-$-polarized pumping laser couples to the transition $|5^2S_{1/2}, F=1\rangle \leftrightarrow |5^2P_{3/2}, F'=1\rangle$ to prepare atoms into $|\uparrow\rangle$ state. The $\sigma^+$-polarized input optical signal field $\hat{\varepsilon}_{in}(z,t)$ is tuned to the transition $|\uparrow\rangle \leftrightarrow |e\rangle$, and goes through cold atoms along $z$ axis. The $\sigma^+$-polarized writing beam $W$ (with a power of ~1 mW and a diameter of 2.5mm) is tuned to the transition $|s\rangle \leftrightarrow |e\rangle$ and goes through the cold atoms with a small angle of ~0.5° from the z axis. The linearly-polarized Raman laser beam (with a ~5.2mm diameter) passing through the cold atoms with an angle of ~1° from the z axis provides $\sigma^+$- and $\sigma^-$-polarized Raman fields $E_{R+}$ and $E_{R-}$, which respectively couple to the transitions $|\uparrow\rangle \leftrightarrow |e\rangle$ and $|\downarrow\rangle \leftrightarrow |e\rangle$ with a detuning $\Delta_R \approx 90MHz$. We use an analogue AOM to modulate Raman laser amplitude and then obtain a Gaussian-shape pulse with a time width of 2.1μs. A $\sigma^+$-polarized reading beam (with a power of ~17.7 mW and a ~2.8mm diameter) couples to the transitions $|s\rangle \leftrightarrow |e\rangle$ to retrieve the two spin-wave components $\hat{S}_{s\uparrow}(z,t)$ and $\hat{S}_{s\downarrow}(z,t)$. In the experiment, the MOT is switched off, and at the same time, a dc magnetic field $B_0$ of ~300 mG along the z axis is applied, so the z-direction quantization axis is well defined. Then the 780 nm pumping and the $\sigma^+$-polarized writing laser beams are turned on to prepare the atoms into the single Zeeman state $|\uparrow\rangle$. After 300 μs, the $\sigma^+$-polarized probe pulse (with a power of ~17μW and a pulse length of 200 ns) is turned on. By switching off the $\sigma^+$-writing laser beams at time $t=0$, the signal field is stored into the atomic ensemble and create a initial spin wave $\hat{S}(z,0) = \langle|\tilde{\rho}_{s\uparrow}(z,0)|\rangle \begin{pmatrix} 1 \\ 0 \end{pmatrix}$. The stored spin wave will be unitarily transformed into $\hat{S}(z,t) = \langle|\tilde{\rho}_{s\uparrow}(z,0)|\rangle \begin{pmatrix} \alpha \\ e^{-i\Delta}\beta \end{pmatrix}$ by Raman manipulation and/or Larmor precession and then is retrieved at time $t$. During the reading process, the two spin-wave components $\hat{S}_{s\uparrow}(z,t)$ and $\hat{S}_{s\downarrow}(z,t)$ are converted into the $\sigma^+$- and $\sigma^-$-polarized output optical signal fields $\varepsilon_+^{out}$ and $\varepsilon_-^{out}$, respectively. After passing through a λ/2-wave-plate, $\varepsilon_+^{out}$ and $\varepsilon_-^{out}$ will become vertical and horizontal polarizations, respectively, and then are split by a polarization splitter (PBS). The separated $\varepsilon_+^{out}$ and $\varepsilon_-^{out}$ signals are detected by D1 and D2 detectors, respectively.

To perform a spin wave $R(\theta,\varphi)$ rotation induced by Raman laser pulse, we first observe the population transfer from $\hat{S}_{s\uparrow}(z,t)$ and $\hat{S}_{s\downarrow}(z,t)$ components as the function of the rotation angle $\theta$. Curves (+) and (-) in Fig.2(a) show the retrieved signals $\varepsilon_+^{out}(t)$ and $\varepsilon_-^{out}(t)$ from the stored spin wave $\tilde{S}(z,t) = \langle|\tilde{\rho}_{s\uparrow}(z,0)|\rangle \begin{pmatrix} 1 \\ 0 \end{pmatrix}$ at $t=17\mu s$ when Raman laser beam is not applied. It can be seen that there is only $\varepsilon_+^{out}$ readout signal (Curve (+)) and no $\varepsilon_-^{out}$ readout signal (Curve (-)), which means that the spin coherence does not freely exchange between $\hat{S}_{s\uparrow}(z,t)$ and $\hat{S}_{s\downarrow}(z,t)$ channels. Curves (+) and (−) in Fig.2(b) exhibit the retrieved signals $\varepsilon_+^{out}$ and $\varepsilon_-^{out}$ simultaneously while a linearly-polarized Raman laser pulse with a peak power $P_R=77mW$ and a pulse width $\tau_R = 2.1\mu s$ is applied. The results show that the spin wave is flipped, which corresponds to a π-pulse operation. To describe the relative value of readout fields $\varepsilon_+^{out}(t)$ and $\varepsilon_-^{out}(t)$ to the original $\varepsilon_0^{out}(t)$ readout, we defined a normalized retrieval efficiency $N_\pm = \dfrac{\int \langle |\varepsilon_\pm^{out}(t)| \rangle^2 dt}{\int \langle |\varepsilon_0^{out}(t)| \rangle^2 dt}$, where $\int \langle |\varepsilon_0^{out}(t)| \rangle^2 dt$ corresponds to the retrieved photon numbers from the spin wave without experiencing Raman laser manipulation at t=17μs. The square (circle) points in Fig.2(c) presents the normalized retrieval efficiencies $N_+$ ($N_-$) as the function of Raman-Rabi frequency $\Omega_R$. The solid curves in Fig.2(c) are the fits to the experimental data based on functions $(1 \pm \cos 2\pi B \Omega_R \tau_R)/2$, with the parameter $B = 0.82$. The fits present sinusoidal Rabi oscillations which are consistent with the theoretical predictions $N_\pm = (1 \pm \cos 2\pi \Omega_R \tau_R)/2$. At $\theta = 2\pi B \Omega_R \tau_R = \pi$ (see upper axis in Fig.2(c)), the population transfer efficiency reaches ~97%.

We then measure the phase $\varphi$ in the rotation $R(\theta, \varphi)$ induced by Raman laser manipulation. The relationship between $\varphi$ and the orientation angle $\varphi_R$ of the linearly-polarized Raman laser is $\varphi = -\pi/2 - 2\varphi_R$. ($\varphi_R = 0$ corresponding to the vertical polarization). By applying a $\pi/2$-Raman-pulse with a variable $\varphi_R$, we transform the initial spin wave $\hat{s}(z,0)$ into $\hat{s}(z,t) = \dfrac{\langle |\tilde{\rho}(z,0)| \rangle}{\sqrt{2}} \begin{pmatrix} 1 \\ e^{i2\varphi_R} \end{pmatrix}$. After a storage time $t=17\mu s$, the spin

wave evolves to $\hat{S}(z,t) = \frac{\langle|\tilde{\rho}_{s\uparrow}(z,0)|\rangle}{\sqrt{2}}\begin{pmatrix}1\\e^{i(2\varphi_R+\phi_{z0})}\end{pmatrix}$ due to Larmor precession in the dc magnetic field $B_0$ and we turn on the $\sigma^+$-polarized reading beam to read the $\hat{s}_{s\uparrow}(z,t)$ and $\hat{s}_{s\downarrow}(z,t)$. A movable half wave plate is inserted front PBS to rotate vertically-polarized $\varepsilon_+^{out}$ and horizontally-polarized $\varepsilon_-^{out}$ fields by $45^0$, respectively. Thus, D1 and D2 detect the fields $\varepsilon_{+45^0}^{out} = (\varepsilon_+^{out} + \varepsilon_-^{out})/\sqrt{2}$ and $\varepsilon_{-45^0}^{out} = (\varepsilon_+^{out} - \varepsilon_-^{out})/\sqrt{2}$ at $\pm 45^0$ orientation, and generate the outputs $\langle|\varepsilon_{\pm 45^0}^{out}|^2\rangle = [\langle|\varepsilon_-^{out} + \varepsilon_+^{out}|^2\rangle]/2 \propto (1 \pm \cos(2\varphi_R + \phi_{z0}))/2$, respectively. Fig.3(a) present the relative retrieved efficiencies $N_{\pm 45^0}$ ($N_{\pm 45^0} = \frac{\int \langle|\varepsilon_{\pm 45^0}^{out}(t)|\rangle^2 dt}{\int \langle|\varepsilon_0^{out}(t)|\rangle^2 dt}$) as the function of $\varphi_R$. The solid lines are fits to the experimental data using functions $(a \pm b\cos(2\varphi_R))/2$, with parameters $\phi_{z0} = 0$, a=0.96 and contrast b=0.96.

By applying a variable field of the magnetic pulse $B(t)$ to the atomic ensemble, we implement spin-wave phase operation i.e. $R_z(\phi_z) = \begin{pmatrix}1 & 0\\0 & e^{i\phi_z}\end{pmatrix}$ rotation. The relative phase shift between $\hat{s}_{s\uparrow}(z,t)$ and $\hat{s}_{s\downarrow}(z,t)$ components induced by Larmor precession in the magnetic field $B=B_0+B(t)$ can be written as $\phi_z = \phi_{z0} + \Delta\phi_z$, where $\phi_{z0}$ is the phase shift induced by the dc magnetic field $B_0$ (~300mG) and $\Delta\phi_z \propto \int B(t)dt$ is the phase shift induced by the pulsed magnetic field $B(t)$. We apply a current pulse $I(t)$ to a pair of coils C1,2 to generate the pulsed magnetic field $B(t)$. The generated pulsed magnetic field $B(t)$ is proportional to $I(t)$. By changing the area $\int I(t)dt$ of current pulse, we can control the changes of $\Delta\phi_z$ and then implement the $R_z(\Delta\phi_z)$ rotation. By observing the interference fringe between the readout fields $\varepsilon_+^{out}(t)$ and $\varepsilon_-^{out}(t)$, we verify the implementation of the $R_z(\Delta\phi_z)$ rotation. In the experiment, we apply a π/2 linearly-polarized Raman laser pulse (with an orientation angle $\varphi_R=8^0$) firstly and then apply a magnetic pulse with a duration of τ~7μs to rotate the initial spin wave $\tilde{s}(z,0)$. After these operations, the spin wave evolves to $\tilde{S}(z,t) = \frac{\langle|\tilde{\rho}_{s\uparrow}(z,0)|\rangle}{\sqrt{2}}\begin{pmatrix} 1 \\ e^{i\Delta_0+i\Delta\phi_z} \end{pmatrix}$, where $\Delta_0 = -\pi/2 - \varphi + \phi_{z0}$ is kept unchanging. At $t$=17μs, we tuned on the reading beam to read the spin wave. Fig.3(b) plot the relative retrieved efficiencies $N_{\pm 45^0}$ (square and circle dots) as the function of the relative magnetic pulse area $A$ [17]. The solid lines are the fits to the measured data $N_{\pm 45^0}$ using sinusoidal functions $(a \pm b\cos 2\pi A)/2$, with parameters $\Delta_0 = 0$, $a$=0.93 and contrast $b$=0.93. The magnetic pulse width is ~7μs in the presented experiment, we believe that it can be further decreased by improving the coils and the current pulse source.

To estimate the π/2-pulse fidelity, we perform a Ramsey experiment by applying two linearly-polarized π/2 Raman laser pulses with a variable time interval $\tau_R$. At $t=17us$, we retrieve the spin wave $\hat{s}(z,t)$ into readout signal fields $\varepsilon_\pm^{out}$. When the movable λ/2 waveplate is removed, the readouts $\varepsilon_\pm^{out}$ signals are injected into D1 and D2, respectively. Fig.4 plots the measured relative photon numbers $N_\pm$ as the function of interval $\tau_R$. The solid lines are the fits to the data $N_\pm$ using the sinusoid function $(1 \pm b_\pm \cos 2\pi\tau_R/T_L)/2$, with the Larmor period $T_L = 2.82\times 10^{-6}s$, and contrasts $b_+$=1 for $N_+$ as well as $b_-$=0.85 for $N_-$. We consider that the deference between $b_+$ and $b_-$ results from the imprecisely Raman laser power for π/2-pulse. According to Rf.[13], we calculate the π/2 fidelity of $F_{\pi/2} = (1+\sqrt{b})/2 \approx 96\%$ by using the lower value of contrast ($b_-$). We believe that $F_{\pi/2}$ can be further improved by more precisely adjusting the Raman laser power to obtain a perfect π/2-pulse operation.

In conclusion, we demonstrated the *R(θ, φ)* and *R$_z$(Δϕ$_z$)* rotations of spin-wave vector on the Bloch-sphere by applying Raman laser and magnetic field pulses. For the case that the initially stored spin wave has one atomic collective excitation (i.e. $N\int \langle \tilde{\rho}_{s\uparrow}(z,0)\tilde{\rho}_{s\uparrow}^{+}(z,0)\rangle dz = 1$), the two spin-wave components sharing the one excitation become two basis states of a qubit. Thus the manipulation demonstrated in the presented work may be used to implement a single qubit gate operation, which has potential applications in QIP based on atomic ensemble.

**Acknowledgement:** We acknowledge funding support from the National Natural Science Foundation of China (No.10874106, 60821004, 10904086), and the 973 Program (2010CB923103). *Corresponding author H. Wang's e-mail address is wanghai@sxu.edu.cn.

17. The relative magnetic pulse area is $A = \int B(t)dt / \int B_0(t)dt$. According to the relation $B(t) \propto I(t)$, we have $A = \int I(t)dt / \int I_0(t)dt$, means we can take the values $A$ by measuring the values $I(t)$. In Fig.3(b), the value $A$ is the area of the current pulse normalized to the current pulse area $\int I_0(t)dt$, where $\int I_0(t)dt$ is the right end of A axis and has been set to 1.

**Figure Caption:**

**Fig. 1** (color online) The experimental scheme for rotations of the spin wave vector. (a-c) atomic level structures of $^{87}$Rb atom for optical storage, Raman manipulation and retrieval. (d) Experimental setup. *R, W* and *P* denote reading, writing, probe laser beams, respectively; BS: beam splitter; C1 and C2: coils; PBS: polarization beam splitter; D1 and D2: photo detectors.

**Fig. 2** (color online) The coherence population transfer between the spin-wave components $\widetilde{S}_{s\uparrow}(z,t)$ and $\widetilde{S}_{s\downarrow}(z,t)$. The curves (+) and (-) in (a) or (b) are the retrieved signal from $\widetilde{S}_{s\uparrow}(z,t)$ and $\widetilde{S}_{s\downarrow}(z,t)$, respectively. (a) is without Raman laser pulse (b) is with π-Raman pulse. The square and circle dots in (c) are the measured Normalized retrieval efficiency $N_+$ and $N_-$ as the function of $\Omega_R$. The solid lines are sinusoidal fits to the data.

**Fig. 3** (color online) The normalized retrieval efficiencies $N_{\pm 45^0}$ as the function of (a) polarization orientation angle $\varphi_R$ of the Raman laser, and (b) magnetic pulse area A. Blue square points: $N_{+45^0}$; red circular points: $N_{-45^0}$. The solid lines are sinusoidal fits to the data.

**Fig. 4** (color online) Ramsey fringes for a pair of π/2 pulses with a variable separate time $\tau_R$. Red circular points (blue square points): the normalized retrieval efficiency $N_+$ ($N_-$). The solid lines are sinusoidal fits to the data.

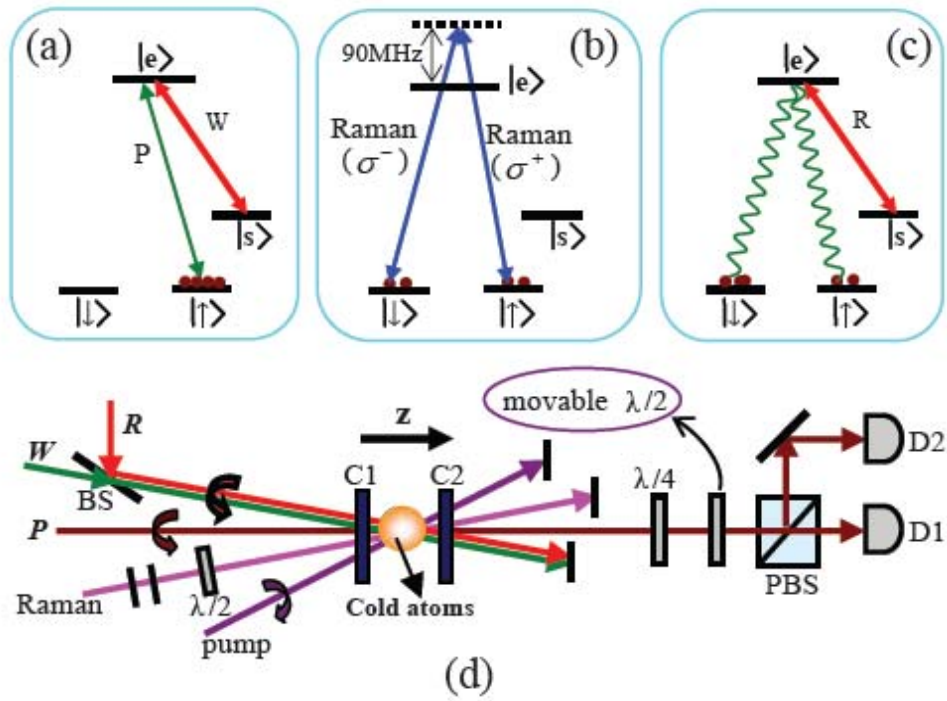

Fig 1

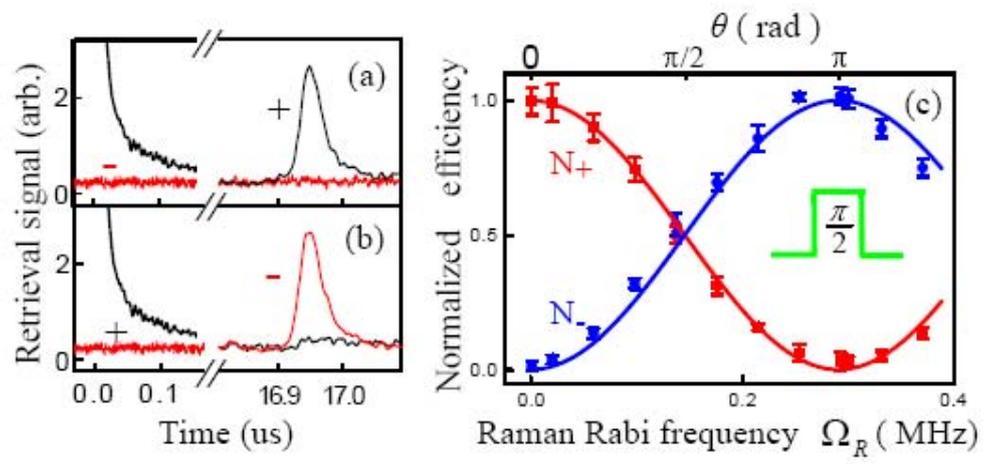

Fig 2

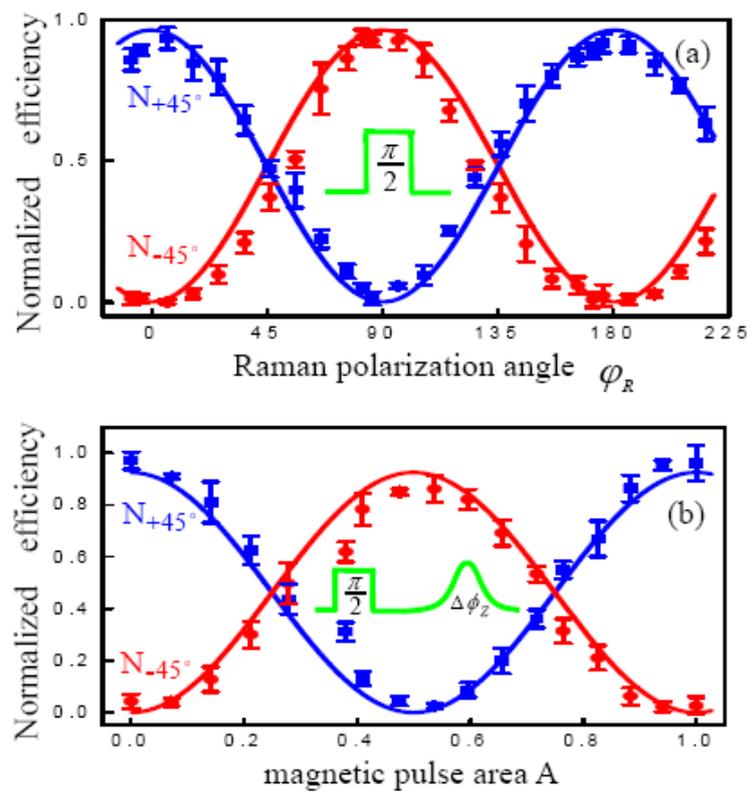

Fig 3

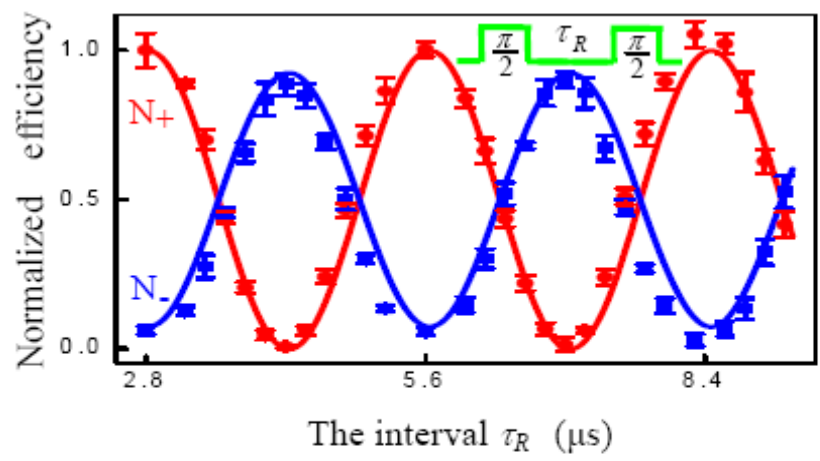

Fig 4